\documentclass[11pt, a4paper]{article}
\usepackage{amsmath, fullpage}
\usepackage{sectsty}
\sectionfont{\fontsize{12}{12}\selectfont}
\subsectionfont{\fontsize{11}{12}\selectfont}
\usepackage{apacite}
\usepackage{amssymb}
\usepackage{graphicx}
\newcommand\sbullet[1][.5]{\mathbin{\vcenter{\hbox{\scalebox{#1}{$\bullet$}}}}}

\usepackage{lipsum}
\usepackage{titlesec}

\usepackage{etoolbox}
\usepackage{authblk}
\AtBeginEnvironment{theorem}{\small}

\usepackage{algorithm}
\usepackage{algpseudocode}

\titlespacing\section{0pt}{12pt plus 4pt minus 2pt}{0pt plus 2pt minus 2pt}
\titlespacing\subsection{0pt}{12pt plus 4pt minus 2pt}{0pt plus 2pt minus 2pt}

\begin{document}
\font\myfont=cmr12 at 15pt
\title{\myfont Electricity Virtual Bidding Strategy Via Entropy-Regularized Stochastic Control Method}
\author[1]{\small Zhou Fang}
\affil[1]{Department of Mathematics, The University of Texas at Austin}
\date{\small Feb 2023}
\maketitle

\begin{abstract}
     We propose a virtual bidding strategy by modeling the price differences between the day-ahead market and the real-time market as Brownian motion with drift, where the drift rate and volatility are functions of meteorological variables. We then transform the virtual bidding problem into a mean-variance portfolio management problem, where we approach the mean-variance portfolio management problem by using the exploratory mean-variance portfolio management framework
    
\end{abstract}

\section{Introduction}
Over the past three decades, the electricity industry has experienced profound changes. Those changes deregulated the electricity markets in most regions, where market participants can buy and sell electricity. Compared to the regulated electricity market, where utility companies control almost everything, a deregulated market has the advantage of being more efficient. However, due to the fact that electricity's demand and supply need to be balanced instantaneously, a deregulated market has more uncertainties. In order to stabilize the electricity market, and to balance demand and supply in real-time, a two-settlement mechanism is introduced in deregulated markets.   

\subsection{Two-Settlement Wholesale Electricity Market}
In each deregulated market, there is one not-for-profit organization called ISO (Independent System Operator) that runs the regional electricity grid. There are seven ISOs in the United States currently including CAISO (California ISO), PJM (Pennsylvania-New Jersey-Maryland Interconnection), and NYISO (New York ISO). Each ISO is responsible for setting electricity prices in its region according to supply and demand. 

Within each deregulated market, there are many places where market participants like power plants or utility companies have access to the electricity grid, where those places are called nodes. The electricity grid transport electricity through transmission lines from one node to another. However, each transmission line has its limit, which is the main reason that different nodes have different electricity prices, which is called locational marginal prices (LMP). In short, LMP at one node is how much unit of electricity cost at that specific node. Next, we will see two-settlement mechanism, which will result a clearer understanding of LMP.   

A two-settlement electricity market consists of day-ahead and real-time markets. The day-ahead market is a forward market and opens one day before the physical delivery or receiving of the electricity. In the day-ahead market, market participants submit bids to ISO. The Bidding information include the volume information that how much electricity they are willing to sell (or buy), the location information that at which node they will deliver (or receive) the electricity, the time information during which hour in the next day they will deliver (or receive) the electricity, and the price information that shows the minimum acceptable price for delivering unit electricity (or maximum price willing to pay for receiving unit electricity). After receiving those bids, ISO will solve a complicated optimization problem to balance the supply and demand over the entire electricity grid with consideration of factors like transmission congestion and cost of losses to get LMP for each node. At each node, a sell bid (buy bid) at that node is accepted if the minimum acceptable price (maximum acceptable price) is lower (higher) than the LMP at that node, and the market participant who submit this bid will receive (charged) credits based on how much electricity she sells (buys) at the rate of LMP. It is worth noting that the day-ahead market's bids promise to deliver or receive electricity for one entire hour. The LMP changes hourly. 

The real-time market acts similarly to the day-ahead market but operates only a few minutes before physical delivery or receiving electricity in order to balance real-time supply and demand. The LMPs change every 5 minutes or 15 minutes depending on the ISO. The real-time market is much more volatile compared to the day-ahead market, and the trading volume is much smaller. In general, more than $98 \%$ of electricity is traded in the day-ahead market.    

\subsection{Virtual Bidding Mechanism}
Some ISOs such as CAISO, PJM, and NYISO allow virtual bidders to participate in the day-ahead and real-time markets. A virtual bidder is a financial entity that doesn't need to own or operate any power-related assets. It can submit bids to buy (sell) a certain amount of electricity on the day-ahead market at a certain node, and sell (buy) the same amount of electricity on the real-time market at the same node. Virtual bidders' profit comes from the price differences between day-ahead and real-time LMP. 

Although virtual bidders, as one type of speculator, are blamed for various reasons, there are several papers that discuss the impact of virtual bidders on the electricity market and show the benefits of virtual bidders for the market. In \cite{ito2016sequential}, it is shown that in a market without virtual bidders, power generators won't bid all of their generation capacity in the day-ahead market in order to increase the day-ahead LMP, and thus make the electricity market more inefficient. In \cite{ito2016sequential}, the author shows that the existence of virtual bidders increases the consumers' welfare, decreases market manipulation by power generators, and increases market efficiency. In addition, \cite{birge2018limits} shows that virtual bidders facing high transaction costs can even increase the prices gap between day-ahead and real-time LMP because even large virtual bidders can suffer losses in the electricity market, but the payoff of holding corresponding transmission right can benefit them more than simply arbitraging in the electricity market.    

In summary, the virtual bidding mechanism is beneficial for the electricity market, and market designers should reduce the obstacles for virtual bidders to trade. 

\subsection{Main Contributions}
There aren't many papers in the literature talking about the virtual bidding strategy mainly because of the complex nature of the such problem. The related papers include \cite{hogan2016virtual}, and \cite{samani2021data}. 

Electricity prices are related to many factors such as meteorological variables, fuel prices, capacities, demands, and grid structure. Some of these factors are unknown like the electricity grid structure, while some of these factors are hard to predict like decisions made by thousands of suppliers and consumers on the market. Although there are papers that recover electricity grid structure from locational marginal prices like \cite{birge2017inverse}, or model electricity prices based on fuel prices \cite{carmona2013electricity}, accurately predicting electricity prices at each node is still an impossible problem because of the stochastic nature of those factors.  

As mentioned above, we only know that electricity prices are related to some factors, but accurately describing this relationship is almost impossible. Due to the stochastic nature of electricity prices, it is natural to approach the virtual bidding problem from a stochastic control approach. In this paper, we model the price differences between the day-ahead market and the real-time market as Brownian motion with drift, and then we develop a mean-variance portfolio management framework to obtain the bidding strategy. The study of the mean-variance starts from the revolutionary paper \cite{selection1952harry}, which solves the portfolio selection problem in a single period. \cite{zhou2000continuous} extends the mean-variance portfolio management problem to a continuous-time setting. Recently, \cite{wang2018exploration} proposes an exploratory mean-variance portfolio management framework in continuous-time, and \cite{wang2019large} extends the exploratory mean-variance framework to the multi-assets case. Since it is generally true that a stochastic policy is more robust than a deterministic policy especially under a stochastic environment, in this paper, we will use the entropy-regularized stochastic control framework that was first proposed in \cite{wang2018exploration} to find an optimal bidding policy for virtual bidders. To our best knowledge, it is the first paper that uses stochastic control tools to get a virtual bidding strategy, which results in a very nice and easy-to-implement bidding strategy.

\section{Model Settings}
The demand for electricity at one node mostly depends on local meteorological variables, the supply at one node depends on the local meteorological variables and fuel prices, and the locational marginal prices additionally depend on the electricity grid structures. Therefore, we consider LMPs as stochastic processes depending on meteorological variables and fuel prices. Since the information on the electricity grid is usually unknown, to capture the impact of the electricity grid's structure on LMPs, the bidding strategy is bidding at many nodes at the same time. In this way, without knowing the exact electricity grid's structure, one can bid with more certainty. 

In the day-ahead market, the LMPs change every hour, and in the real-time market, the LMPs change every 5 or 15 minutes depending on the ISO. Since virtual bidders, need to submit bids in the day-ahead market for a specific hour $h$ in the next day and liquidate their position for that hour $h$ in the real-time market, we will assume that the LMPs at which a virtual bidder liquidates her position on the real-time market is the average of LMPs for the hour $h$ in the real-time market. In the following discussion, the setting is restricted to a fixed hour $h$, where $h$ can be any one of 24 hours in a day. The subscript $t$ in the following notation indicates the day. 

Assume there are $n$ nodes where virtual bidders can submit bids, denote as $X = \{x_1, x_2, ..., x_n\}$. Let $\mathbf{\theta_t= (\theta_t^{1}, \theta_t^{2}, ..., \theta_t^{n})}$ be the vector of meteorological variables at time $t$, where each $\mathbf{\theta_t^{i}} = (\theta_t^{i,1}, \theta_t^{i,2}, ..., \theta_t^{i,k})$ is the vector of $k$ meteorological variables at node $i$. Possible meteorological variables to consider include wind speed, solar radiance, temperature, humidity, and so on, we will discuss them in more detail in empirical studies. Since intraday meteorological variables forecast can be accurate, we assume that the meteorological variables information of the current day is available for virtual bidders when they submit bids on the day-ahead market. In this paper, we won't consider using any forecast model to forecast the meteorological variables for the next day. The reason is, because of the lack of professional equipment and resources, and the historic forecast meteorological variables data aren't published by professional institutes, the best forecast model we could use is the time series model. However, time series models aren't that accurate, and time series models are also Markovian, which means they depend on the current meteorological variables, therefore, the forecast model is also not needed if the bidding strategy is based on the current day's meteorological variables. We do believe that having a sophisticated forecast model and keeping forecast data in the records is necessary to get a better bidding strategy and will show in the empirical study section the difference between the bidding strategy only assuming the current day's meteorological variables and the bidding strategy assuming the next day's realized meteorological variables as the forecast meteorological variables (in other words, assume the forecast can be extremely accurate). So, $\boldsymbol \theta_t$ can either represent the current day's meteorological variables or represent the current day's meteorological variables and the forecast meteorological variables for the next day, which makes our model quite flexible.   

Let $\mathbf{f_t}$ be a $n-$dimensional vector that represents the difference of log LMPs on the day-ahead at time $t$ and log LMPs on the real-time market on the day $t+24$ (we use 1 hour as the basic unit here) at the above mentioned $n$ nodes. Since LMPs depend on meteorological variables, and fuel prices, we can write $\mathbf{f_t = f_t(\theta_t)}$. Since the electricity grid's conditions can change over time, and supply and demand can't be the same even under the same weather conditions, then it is reasonable to model the difference between day-ahead and real-time log LMPs as random variables. We assume $\mathbf{f_t(\theta_t)}$ have the following form, 
$$\mathbf{f_t} = \mathbf{b_t(\boldsymbol \theta_t)} + \boldsymbol \sigma_t (\boldsymbol \theta_t) \Delta \boldsymbol W_t $$
where $\boldsymbol \sigma_t(\boldsymbol \theta_t) = (\sigma^{ij}_t(\boldsymbol \theta_t))$ is a $n \times n$ matrix, and $\Delta \boldsymbol W_t$ is n-dimensional independent Gaussian distribution. It is worth noting that the unit time here is one hour.     

As mentioned above, to make the bidding strategy more robust, we use a stochastic policy instead of a deterministic one, and also consider the meteorological variables and fuel prices. More specifically, let $\boldsymbol{\pi}$ denote our policy, given current total wealth, meteorological variables and fuel prices $(X_t, \boldsymbol \theta_t)$, then $\boldsymbol{\pi}(q_t| X_t, \boldsymbol \theta_t)$ denotes the probability density that the virtual bidder will distribute her wealth in the following way $-$ $\mathbf{q_t} = (q_t^1, q_t^2, ..., q_t^n)$ meaning put $q_t^i$ dollars to buy (sell if $q_t^i$ is negative) at node $i$ in the day-ahead market. Notice that because of the uniform auction nature of the day-ahead market, virtual bidders may face difficulty to achieve the precise distribution of their wealth as planned. However, since LMPs of the day-ahead market is more predictable, in \cite{birge2018limits}, it shows a regression between previous day-ahead market LMPs, and current day-ahead market LMPs have an R-squared larger than 0.8, which means it is possible to distribute wealth roughly the way as planned. Some statistical methods need to be used to let the distribution of wealth as close to the plan as possible, but that's beyond the scope of this paper, and guarantees separate research. In this paper, we just assume that we can distribute wealth as planned.  

In addition, to make our virtual bids successfully executed and not affect locational marginal prices with those virtual bids, we assume our buy bids are at 0, and our sell bids are at high prices, while the volume of our bids at each node will be small compare to the trading volume at that node.

\section{Entropy-Regularized Stochastic Control}
Consider the classical mean-variance portfolio problem. Let's fix $z$ to be the expected return, and we would like to minimize the variance. Denote $X_t$ as our wealth process, and the terminal time is $T$, then we have an optimization problem
\begin{align*}
    \text{min Var}[X_T] \\
    \text{s.t. } \mathbb{E}[X_T] = z
\end{align*}
Let the Lagrangian multiplier be $w$, and by the fact that $\text{Var}[X_T] = \mathbb{E}[X_T^2] - z^2$, we transform the above optimization problem into the following form,
$$\text{min } \mathbb{E}[X_T^2] - z^2 - 2w(\mathbb{E}[X_T] - z) = \text{min } \mathbb{E}[(X_T - w)^2] - (w - z)^2$$
From the above derivation, for no matter what stochastic process $X_T$, to minimize the above objective, we need to know $\mathbb{E}[X_T]$ and $\mathbb{E}[X_T^2]$. 

Proposed in \cite{wang2018exploration}, the entropy-regularized stochastic control method uses stochastic policy. Since the virtual bidding mechanism is in discrete time setting, which increases the difficulty. Because usually under discrete-time setting, it is hard to extract interesting information from value function, which also increases the difficulty to write policy explicitly. However, the terminal time $T$ we consider in this case is usually a year, then it is reasonable to consider the problem first under continuous time setting, and we can discretize the time when implementing and testing our strategy. 

Denote the total wealth at day $t$ as $X_t$. To simplify the notations in the following computation, let $\boldsymbol b_t = \boldsymbol b_t(\boldsymbol \theta_t)$, and $\boldsymbol \sigma_t = \boldsymbol \sigma_t(\boldsymbol \theta_t)$, assume the holding at time $t$ is $\boldsymbol q_t$, then the change of wealth will be
$$d X_t =  \boldsymbol q_t^{T} (\mathbf{b_t} d t + \boldsymbol \sigma_t d \boldsymbol W_t)$$

Now, since the policy is stochastic, it is hard to write the wealth process' dynamics as easy as the above one. Denote the dynamics of wealth under control $\boldsymbol \pi$ as $d X_t^{\boldsymbol \pi}$. To do the mean-variance optimization, we need to know $\mathbb{E}[X_t^{\boldsymbol \pi}]$, and $\mathbb{E}[(X_t^{\boldsymbol \pi})^2 ]$. Although it is hard to know the exact form of $d X_t^{\boldsymbol \pi}$, we can still manage to know the $\mathbb{E}[d  X_t^{\boldsymbol \pi}]$, and $\mathbb{E}[(d  X_t^{\boldsymbol \pi})^2 ]$. Assume we sample $d X_t^{\boldsymbol \pi}$ for $M$ times, and the stochastic policy is independent of Brownian motion, then, by the law of large number, we have the following formulas
\begin{align*}
    \sum_{i = 1}^{M} d X_t^{\boldsymbol \pi, i} &= \sum_{i = 1}^{M} (\boldsymbol q_t^i)^T \boldsymbol b_t dt + (\boldsymbol q_t^i)^T \boldsymbol \sigma_t d \boldsymbol W_t \\
    &\rightarrow \mathbb{E}[ \boldsymbol b_t^T \Big( \int_{\mathbb{R}^n} \boldsymbol q_t \boldsymbol \pi(\boldsymbol q_t) d \boldsymbol q_t \Big) dt + \Big( \int_{\mathbb{R}^n} \boldsymbol q_t^T \boldsymbol \sigma_t \boldsymbol \pi (\boldsymbol q_t) d \boldsymbol q_t \Big) d \boldsymbol W_t ] \\
    &=  \boldsymbol b_t^T \Big( \int_{\mathbb{R}^n} \boldsymbol q_t \boldsymbol \pi(\boldsymbol q_t) d \boldsymbol q_t \Big) dt \\
    &= \mathbb{E}[d X_t^{\boldsymbol \pi}]            \\
    \sum_{i = 1}^{M} (d X_t^{\boldsymbol \pi, i})^2 &= \sum_{i = 1}^{M} (\boldsymbol q_t^{i})^T \boldsymbol \sigma_t \boldsymbol \sigma_t^T \boldsymbol q_t^{i} dt \\ &\rightarrow \mathbb{E}[ \Big( \int_{\mathbb{R}^n} (\boldsymbol q_t)^T \boldsymbol \sigma_t \boldsymbol \sigma_t^T \boldsymbol q_t \boldsymbol \pi(\boldsymbol q_t) d \boldsymbol q_t \Big) dt] \\
    &= \Big( \int_{\mathbb{R}^n} (\boldsymbol q_t)^T \boldsymbol \sigma_t \boldsymbol \sigma_t^T \boldsymbol q_t \boldsymbol \pi(\boldsymbol q_t) d \boldsymbol q_t \Big) dt \\
    &= \mathbb{E}[(d X_t^{\boldsymbol \pi})^2]
\end{align*}

The above results are good enough to get the optimal stochastic policy. Fix $z$, which is the expected return, and $w$ is the fixed Lagrangian multiplier. If we also take the degree of exploration of the stochastic policy into consideration at each time, then the mean-variance optimization problem becomes the following, 
$$
\underset{\boldsymbol \pi}{min} \hspace{0.1cm} \mathbb{E}[(X_T^{\boldsymbol \pi} - w)^2 + \gamma \int_{0}^{T} \int_{\mathbb{R}^n} \boldsymbol \pi(\boldsymbol q_t) \text{log} \boldsymbol \pi(\boldsymbol q_t) d \boldsymbol q_t] - (w - z)^2
$$
if the current wealth is $X_t$, and current policy is $\boldsymbol \pi$, denote the value function as
$$J(X_t, \boldsymbol \pi) = \mathbb{E}[(X_T^{\boldsymbol \pi} - w)^2 + \gamma \int_{t}^{T} \int_{\mathbb{R}^n} \boldsymbol \pi(\boldsymbol q_t) \text{log} \boldsymbol \pi(\boldsymbol q_t) d \boldsymbol q_t | X_t] - (w - z)^2$$
then, by the dynamic programming principle, we have 
$$
 \mathbb{E}[d X_t^{\boldsymbol \pi}] \Delta_x J(X_t, \boldsymbol \pi) + \Delta_t J(X_t, \boldsymbol \pi) + \frac{1}{2}\mathbb{E}[(d X_t^{\boldsymbol \pi})^2] \Delta_{xx} J(X_t, \boldsymbol \pi) + \int_{\mathbb{R}^n} \boldsymbol \pi(\boldsymbol q_t) \text{log} \boldsymbol \pi(\boldsymbol q_t) d \boldsymbol q_t = 0
$$
since the true value function is
$$V(X_t) = \underset{\boldsymbol \pi}{\text{min}} J(X_t, \boldsymbol \pi)$$ 
the HJB equation is
$$
\Delta_t V(X_t) + \underset{\boldsymbol \pi}{\text{min}} \Big\{ \int_{\mathbb{R}^n} \bigg( \boldsymbol b_t^T \boldsymbol q_t \Delta_x V(X_t) + \frac{1}{2} (\boldsymbol q_t)^T \boldsymbol \sigma_t \boldsymbol \sigma_t^T \boldsymbol q_t \Delta_{xx} V(X_t) + \gamma \text{log} \boldsymbol \pi(\boldsymbol q_t) \bigg) \boldsymbol \pi(\boldsymbol q_t) d \boldsymbol q_t    \Big\}
$$
to find the candidate solution for the above HJB equation, we first need to minimize the following formula, 
$$\int_{\mathbb{R}^n} \bigg( \boldsymbol b_t^T \boldsymbol q_t \Delta_x V(X_t) + \frac{1}{2} (\boldsymbol q_t)^T \boldsymbol \sigma_t \boldsymbol \sigma_t^T \boldsymbol q_t \Delta_{xx} V(X_t) + \gamma \text{log} \boldsymbol \pi(\boldsymbol q_t) \bigg) \boldsymbol \pi(\boldsymbol q_t) d \boldsymbol q_t$$
through simple observations, we know that the following quantity should be constant, because if it is not constant, then we can easily improve the current policy by increasing the density where the following quantity is smaller while doing the opposite where the following quantity is larger. 
$$ \boldsymbol b_t^T \boldsymbol q_t \Delta_x V(X_t) + \frac{1}{2} (\boldsymbol q_t)^T \boldsymbol \sigma_t \boldsymbol \sigma_t^T \boldsymbol q_t \Delta_{xx} V(X_t) + \gamma \text{log} \boldsymbol \pi(\boldsymbol q_t) = C$$
a straight computation will show 
$$\boldsymbol \pi(\boldsymbol q_t) = C e^{-\frac{1}{\gamma} \boldsymbol b_t^T \boldsymbol q_t \Delta_x V(X_t) + \frac{1}{2} (\boldsymbol q_t)^T \boldsymbol \sigma_t \boldsymbol \sigma_t^T \boldsymbol q_t \Delta_{xx} V(X_t)  }   $$
also by the fact that $\boldsymbol \pi$ is a probability density distribution, then we have $$\boldsymbol \pi(\boldsymbol q_t) = \frac{e^{-\frac{1}{\gamma} \big[ \boldsymbol b_t^T \boldsymbol q_t \Delta_x V(X_t) + \frac{1}{2} (\boldsymbol q_t)^T \boldsymbol \sigma_t \boldsymbol \sigma_t^T \boldsymbol q_t \Delta_{xx} V(X_t) \big] }}{\int_{\mathbb{R}^n} e^{-\frac{1}{\gamma} \big [ \boldsymbol b_t^T \boldsymbol q_t \Delta_x V(X_t) + \frac{1}{2} (\boldsymbol q_t)^T \boldsymbol \sigma_t \boldsymbol \sigma_t^T \boldsymbol q_t \Delta_{xx} V(X_t) \big ]  } d \boldsymbol q_t}$$
Therefore, we know the stochastic policy is Gaussian distribution
$$\boldsymbol \pi ( \hspace{0.1cm} \sbullet \hspace{0.1cm} | X_t) \sim \mathcal{N}\Big( - (\boldsymbol \sigma_t \boldsymbol \sigma_t^{T})^{-1} \boldsymbol b_t\frac{\Delta_x V(X_t)}{\Delta_{xx} V(X_t)} \hspace{0.1cm}, \hspace{0.1cm}  (\boldsymbol \sigma_t \boldsymbol \sigma_t^{T})^{-1} \frac{\gamma}{\Delta_{xx} V(X_t)} \Big )$$
Then, we can get analytic solution for $V(X_t)$ and $\boldsymbol \pi (\boldsymbol q_t)$. The analytic solution for the policy is, 
$$
\boldsymbol \pi ( \hspace{0.1cm} \sbullet \hspace{0.1cm} | X_t) \sim \mathcal{N} \Big( - (\boldsymbol \sigma_t \boldsymbol \sigma_t^{T})^{-1} \boldsymbol b_t (X_t - \frac{z e^{\boldsymbol b_t^T (\boldsymbol \sigma_t \boldsymbol \sigma_t^{T})^{-1} \boldsymbol b_t T} - X_0}{e^{\boldsymbol b_t^T (\boldsymbol \sigma_t \boldsymbol \sigma_t^{T})^{-1} \boldsymbol b_t T} - 1}) \hspace{0.1cm}, \hspace{0.1cm}  \frac{\gamma}{2} (\boldsymbol \sigma_t \boldsymbol \sigma_t^{T})^{-1} e^{\boldsymbol b_t^T (\boldsymbol \sigma_t \boldsymbol \sigma_t^{T})^{-1} \boldsymbol b_t (T - t)}  \Big )
$$

\section{Estimation of Model Parameters}
The next step is estimating the parameters for $f_t(\boldsymbol \theta_t)$. Under current setting, the inputs are $\boldsymbol \theta_t$, and we wish the outputs to be $\boldsymbol b_t(\boldsymbol \theta_t)$, and $\boldsymbol \sigma_t(\boldsymbol \theta_t)$. Assume we already have a parametric form for those two coefficients, denote the parameter as $\phi$, and the estimated coefficients as $\boldsymbol b^{\phi}_t(\boldsymbol \theta_t)$, and $\boldsymbol \sigma^{\phi}_t(\boldsymbol \theta_t)$. Denote the resulting multivariate normal distribution as $\mathbf{f^{\phi}_t( \boldsymbol \theta_t)}$. 

Assume we have past electricity prices, $\mathbf{f}_t = \mathbf{f}_t(\boldsymbol \theta_t)$, where $1 \leq t \leq T$. To accurately estimate the parameters, One popular method is the maximum likelihood estimation. Thus, the object is to maximize the following,
$$\underset{\phi}{\text{max}} \hspace{0.1cm} \underset{1 \leq t \leq T}{\sum}  \text{log} \frac{1}{\sqrt{(2\pi)^{n} det(\boldsymbol \sigma^{\phi}_t)}} e^{-\frac{1}{2} (\boldsymbol f_t - \boldsymbol b^{\phi}_t)^{T} (\boldsymbol \sigma^{\phi}_t)^{-1} (\boldsymbol f_t - \boldsymbol b^{\phi}_t)}  $$
denote the objective function as $\mathcal{H}(\phi)$,
\begin{align*}
\mathcal{H}(\phi) &=   \underset{1 \leq t \leq T}{\sum}  \text{log} \frac{1}{\sqrt{(2\pi)^{n} det(\boldsymbol \sigma^{\phi}_t)}} e^{-\frac{1}{2} (\boldsymbol f_t - \boldsymbol b^{\phi}_t)^{T} (\boldsymbol \sigma^{\phi}_t)^{-1} (\boldsymbol f_t - \boldsymbol b^{\phi}_t)}  \\ 
&= \underset{1 \leq t \leq T}{\sum} -\frac{n}{2} \text{log} (2 \pi) - \frac{1}{2} \mathrm{log} (\mathrm{det}(\boldsymbol \sigma^{\phi}_t)) - \frac{1}{2} (\boldsymbol f_t - \boldsymbol b^{\phi}_t)^{T} (\boldsymbol \sigma^{\phi}_t)^{-1} (\boldsymbol f_t - \boldsymbol b^{\phi}_t)
\end{align*}
let $A(\phi)$ be a matrix, $\frac{\partial}{\partial \phi} \text{det} A(\phi)  = \text{det} A \hspace{0.1cm} \text{tr}(A^{-1} \frac{A(\phi)}{\partial \phi})$, and $\frac{\partial A(\phi)^{-1}}{\partial \phi} = - A^{-1} \frac{\partial A(\phi)}{\partial \phi} A^{-1}$. Then the partial derivative of $\mathcal{H}(\phi)$ is,

\begin{align*}
\frac{\partial \mathcal{H}(\phi)}{\partial \phi} &=   \underset{1 \leq t \leq T}{\sum} -\frac{1}{2} \text{tr}\big((\boldsymbol \sigma^{\phi}_t)^{-1} \frac{\partial \boldsymbol \sigma^\phi_t}{\partial \phi}\big ) + \Big (\frac{\partial \boldsymbol b^{\phi}_t }{\partial \phi} \Big )^{T} (\boldsymbol \sigma^{\phi}_t)^{-1} (\boldsymbol f_t - \boldsymbol b^{\phi}_t) - \frac{1}{2}(\boldsymbol f_t - \boldsymbol b^{\phi}_t)^{T} \frac{\partial (\boldsymbol \sigma^{\phi}_t)^{-1}}{\partial \phi}(\boldsymbol f_t - \boldsymbol b^\phi_t)
\\
&=   \underset{1 \leq t \leq T}{\sum} -\frac{1}{2} \text{tr}\big((\boldsymbol \sigma^{\phi}_t)^{-1} \frac{\partial \boldsymbol \sigma^\phi_t}{\partial \phi}\big ) + \Big (\frac{\partial \boldsymbol b^{\phi}_t }{\partial \phi} \Big )^{T} (\boldsymbol \sigma^{\phi}_t)^{-1} (\boldsymbol f_t - \boldsymbol b^{\phi}_t) + \frac{1}{2}(\boldsymbol f_t - \boldsymbol b^\phi_t)^{T}(\boldsymbol \sigma^\phi_t)^{-1}\frac{\partial \boldsymbol \sigma^\phi_t}{\partial \phi} (\boldsymbol \sigma^\phi_t)^{-1} (\boldsymbol f_t - \boldsymbol b^\phi_t) 
\end{align*}
thus, one can update the parameters, 
\begin{align*}
\tilde{\phi}  &=  \phi + \alpha \frac{\partial \mathcal{H}(\phi)}{\partial \phi} 
\\
&= \phi + \alpha   \underset{1 \leq t \leq T}{\sum} -\frac{1}{2} \text{tr}\big((\boldsymbol \sigma^{\phi}_t)^{-1} \frac{\partial \boldsymbol \sigma^\phi_t}{\partial \phi}\big ) + \Big (\frac{\partial \boldsymbol b^{\phi}_t }{\partial \phi} \Big )^{T} (\boldsymbol \sigma^{\phi}_t)^{-1} (\boldsymbol f_t - \boldsymbol b^{\phi}_t) + \frac{1}{2}(\boldsymbol f_t - \boldsymbol b^\phi_t)^{T}(\boldsymbol \sigma^\phi_t)^{-1}\frac{\partial \boldsymbol \sigma^\phi_t}{\partial \phi} (\boldsymbol \sigma^\phi_t)^{-1} (\boldsymbol f_t - \boldsymbol b^\phi_t)  
\end{align*}

This gives us a general framework to estimate parameters, with which one is able to use a more complicated model involving fancy neural networks. In this paper, to keep things simple, and tractable, we only parametrize $\mathbf{b_t^{\phi}}$ by using multiple linear regression and estimate the covariance matrix $\boldsymbol \sigma_t$ based on the past 60 days' prices differences. 

\section{Empirical Studies}
During the empirical study, we selected a value of $z = 105$ and $\gamma = 0.001$ for the strategy. The initial wealth is $100$. We randomly picked 9 nodes in the CAISO system and tested the strategy over three different time periods, which represented different seasons: January 2022 (winter), August 2022 (summer), and October 2022 (fall). We excluded data from Jan 5, Aug 1, 24, 25, and Oct 17 due to incomplete price information.

As mentioned in section 5, in this paper we will only estimate parameters under simple settings, where the drift rate at each location is a linear function of three meteorological variables, which are temperature, humidity, and wind speed.  We also assume the covariance matrix is constant. This basic model doesn't have good prediction ability, which causes the parameters in the holdings/control achieved to a ridiculous level and results in an unrealistic wealth process. However, when perturbating the true parameters by noise, the performance is quite good, which means that with a good estimation of parameters, this framework yields a robust strategy. Figures 1 -- 3 are the backtesting performance in January, August, and October when the estimation of parameters is accurate (with noise added). 

\section{Conclusion}
In this paper, we consider virtual bidding in a mean-variance framework by modeling price differences between the day-ahead market and the real-time market as Brownian motion with drift. The numerical results indicate that with an accurate estimation of parameters, the framework yields a profitable and robust strategy. Future work could explore more complex models and improved estimation techniques to further improve the performance of this framework.

\newpage
\bibliography{Reference.bib}

\begin{thebibliography}{}

\bibitem [\protect \citeauthoryear {%
Birge%
, Horta{\c{c}}su%
, Mercadal%
\BCBL {}\ \BBA {} Pavlin%
}{%
Birge%
\ \protect \BOthers {.}}{%
{\protect \APACyear {2018}}%
}]{%
birge2018limits}
\APACinsertmetastar {%
birge2018limits}%
\begin{APACrefauthors}%
Birge, J\BPBI R.%
, Horta{\c{c}}su, A.%
, Mercadal, I.%
\BCBL {}\ \BBA {} Pavlin, J\BPBI M.%
\end{APACrefauthors}%
\unskip\
\newblock
\APACrefYearMonthDay{2018}{}{}.
\newblock
{\BBOQ}\APACrefatitle {Limits to Arbitrage in Electricity Markets: A case study
  of MISO} {Limits to arbitrage in electricity markets: A case study of
  miso}.{\BBCQ}
\newblock
\APACjournalVolNumPages{Energy Economics}{75}{}{518--533}.
\PrintBackRefs{\CurrentBib}

\bibitem [\protect \citeauthoryear {%
Birge%
, Horta{\c{c}}su%
\BCBL {}\ \BBA {} Pavlin%
}{%
Birge%
\ \protect \BOthers {.}}{%
{\protect \APACyear {2017}}%
}]{%
birge2017inverse}
\APACinsertmetastar {%
birge2017inverse}%
\begin{APACrefauthors}%
Birge, J\BPBI R.%
, Horta{\c{c}}su, A.%
\BCBL {}\ \BBA {} Pavlin, J\BPBI M.%
\end{APACrefauthors}%
\unskip\
\newblock
\APACrefYearMonthDay{2017}{}{}.
\newblock
{\BBOQ}\APACrefatitle {Inverse optimization for the recovery of market
  structure from market outcomes: An application to the MISO electricity
  market} {Inverse optimization for the recovery of market structure from
  market outcomes: An application to the miso electricity market}.{\BBCQ}
\newblock
\APACjournalVolNumPages{Operations Research}{65}{4}{837--855}.
\PrintBackRefs{\CurrentBib}

\bibitem [\protect \citeauthoryear {%
Carmona%
, Coulon%
\BCBL {}\ \BBA {} Schwarz%
}{%
Carmona%
\ \protect \BOthers {.}}{%
{\protect \APACyear {2013}}%
}]{%
carmona2013electricity}
\APACinsertmetastar {%
carmona2013electricity}%
\begin{APACrefauthors}%
Carmona, R.%
, Coulon, M.%
\BCBL {}\ \BBA {} Schwarz, D.%
\end{APACrefauthors}%
\unskip\
\newblock
\APACrefYearMonthDay{2013}{}{}.
\newblock
{\BBOQ}\APACrefatitle {Electricity price modeling and asset valuation: a
  multi-fuel structural approach} {Electricity price modeling and asset
  valuation: a multi-fuel structural approach}.{\BBCQ}
\newblock
\APACjournalVolNumPages{Mathematics and Financial Economics}{7}{2}{167--202}.
\PrintBackRefs{\CurrentBib}

\bibitem [\protect \citeauthoryear {%
Hogan%
}{%
Hogan%
}{%
{\protect \APACyear {2016}}%
}]{%
hogan2016virtual}
\APACinsertmetastar {%
hogan2016virtual}%
\begin{APACrefauthors}%
Hogan, W\BPBI W.%
\end{APACrefauthors}%
\unskip\
\newblock
\APACrefYearMonthDay{2016}{}{}.
\newblock
{\BBOQ}\APACrefatitle {Virtual bidding and electricity market design} {Virtual
  bidding and electricity market design}.{\BBCQ}
\newblock
\APACjournalVolNumPages{The Electricity Journal}{29}{5}{33--47}.
\PrintBackRefs{\CurrentBib}

\bibitem [\protect \citeauthoryear {%
Ito%
\ \BBA {} Reguant%
}{%
Ito%
\ \BBA {} Reguant%
}{%
{\protect \APACyear {2016}}%
}]{%
ito2016sequential}
\APACinsertmetastar {%
ito2016sequential}%
\begin{APACrefauthors}%
Ito, K.%
\BCBT {}\ \BBA {} Reguant, M.%
\end{APACrefauthors}%
\unskip\
\newblock
\APACrefYearMonthDay{2016}{}{}.
\newblock
{\BBOQ}\APACrefatitle {Sequential markets, market power, and arbitrage}
  {Sequential markets, market power, and arbitrage}.{\BBCQ}
\newblock
\APACjournalVolNumPages{American Economic Review}{106}{7}{1921--57}.
\PrintBackRefs{\CurrentBib}

\bibitem [\protect \citeauthoryear {%
Samani%
, Kohansal%
\BCBL {}\ \BBA {} Mohsenian-Rad%
}{%
Samani%
\ \protect \BOthers {.}}{%
{\protect \APACyear {2021}}%
}]{%
samani2021data}
\APACinsertmetastar {%
samani2021data}%
\begin{APACrefauthors}%
Samani, E.%
, Kohansal, M.%
\BCBL {}\ \BBA {} Mohsenian-Rad, H.%
\end{APACrefauthors}%
\unskip\
\newblock
\APACrefYearMonthDay{2021}{}{}.
\newblock
{\BBOQ}\APACrefatitle {A Data-Driven Convergence Bidding Strategy Based on
  Reverse Engineering of Market Participants’ Performance: A Case of
  California ISO} {A data-driven convergence bidding strategy based on reverse
  engineering of market participants’ performance: A case of california
  iso}.{\BBCQ}
\newblock
\APACjournalVolNumPages{IEEE Transactions on Power Systems}{37}{3}{2122--2136}.
\PrintBackRefs{\CurrentBib}

\bibitem [\protect \citeauthoryear {%
Selection%
}{%
Selection%
}{%
{\protect \APACyear {1952}}%
}]{%
selection1952harry}
\APACinsertmetastar {%
selection1952harry}%
\begin{APACrefauthors}%
Selection, P.%
\end{APACrefauthors}%
\unskip\
\newblock
\APACrefYearMonthDay{1952}{}{}.
\newblock
{\BBOQ}\APACrefatitle {Harry Markowitz} {Harry markowitz}.{\BBCQ}
\newblock
\APACjournalVolNumPages{The Journal of Finance}{7}{1}{77--91}.
\PrintBackRefs{\CurrentBib}

\bibitem [\protect \citeauthoryear {%
Wang%
}{%
Wang%
}{%
{\protect \APACyear {2019}}%
}]{%
wang2019large}
\APACinsertmetastar {%
wang2019large}%
\begin{APACrefauthors}%
Wang, H.%
\end{APACrefauthors}%
\unskip\
\newblock
\APACrefYearMonthDay{2019}{}{}.
\newblock
{\BBOQ}\APACrefatitle {Large scale continuous-time mean-variance portfolio
  allocation via reinforcement learning} {Large scale continuous-time
  mean-variance portfolio allocation via reinforcement learning}.{\BBCQ}
\newblock
\APACjournalVolNumPages{arXiv preprint arXiv:1907.11718}{}{}{}.
\PrintBackRefs{\CurrentBib}

\bibitem [\protect \citeauthoryear {%
Wang%
, Zariphopoulou%
\BCBL {}\ \BBA {} Zhou%
}{%
Wang%
\ \protect \BOthers {.}}{%
{\protect \APACyear {2018}}%
}]{%
wang2018exploration}
\APACinsertmetastar {%
wang2018exploration}%
\begin{APACrefauthors}%
Wang, H.%
, Zariphopoulou, T.%
\BCBL {}\ \BBA {} Zhou, X.%
\end{APACrefauthors}%
\unskip\
\newblock
\APACrefYearMonthDay{2018}{}{}.
\newblock
{\BBOQ}\APACrefatitle {Exploration versus exploitation in reinforcement
  learning: a stochastic control approach} {Exploration versus exploitation in
  reinforcement learning: a stochastic control approach}.{\BBCQ}
\newblock
\APACjournalVolNumPages{arXiv preprint arXiv:1812.01552}{}{}{}.
\PrintBackRefs{\CurrentBib}

\bibitem [\protect \citeauthoryear {%
Zhou%
\ \BBA {} Li%
}{%
Zhou%
\ \BBA {} Li%
}{%
{\protect \APACyear {2000}}%
}]{%
zhou2000continuous}
\APACinsertmetastar {%
zhou2000continuous}%
\begin{APACrefauthors}%
Zhou, X\BPBI Y.%
\BCBT {}\ \BBA {} Li, D.%
\end{APACrefauthors}%
\unskip\
\newblock
\APACrefYearMonthDay{2000}{}{}.
\newblock
{\BBOQ}\APACrefatitle {Continuous-time mean-variance portfolio selection: A
  stochastic LQ framework} {Continuous-time mean-variance portfolio selection:
  A stochastic lq framework}.{\BBCQ}
\newblock
\APACjournalVolNumPages{Applied Mathematics and Optimization}{42}{}{19--33}.
\PrintBackRefs{\CurrentBib}

\end{thebibliography}
\bibliographystyle{apacite}

\newpage

\begin{figure}[p]
    \centering
    \includegraphics[width = \textwidth]{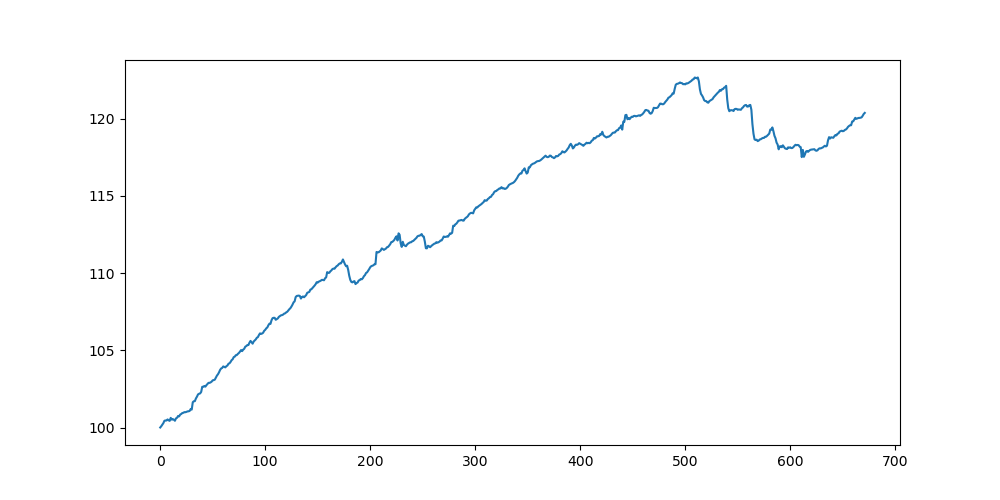}
    \caption{January Returns Example 2}
    \label{fig:my_label}
\end{figure}

\begin{figure}[p]
    \centering
    \includegraphics[width = \textwidth]{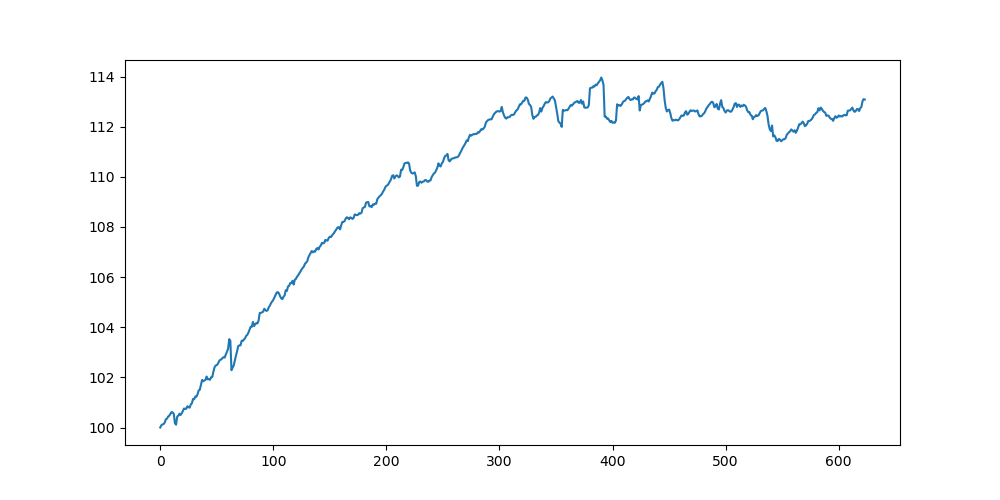}
    \caption{August Returns Example 2}
    \label{fig:my_label}
\end{figure}

\begin{figure}[p]
    \centering
    \includegraphics[width = \textwidth]{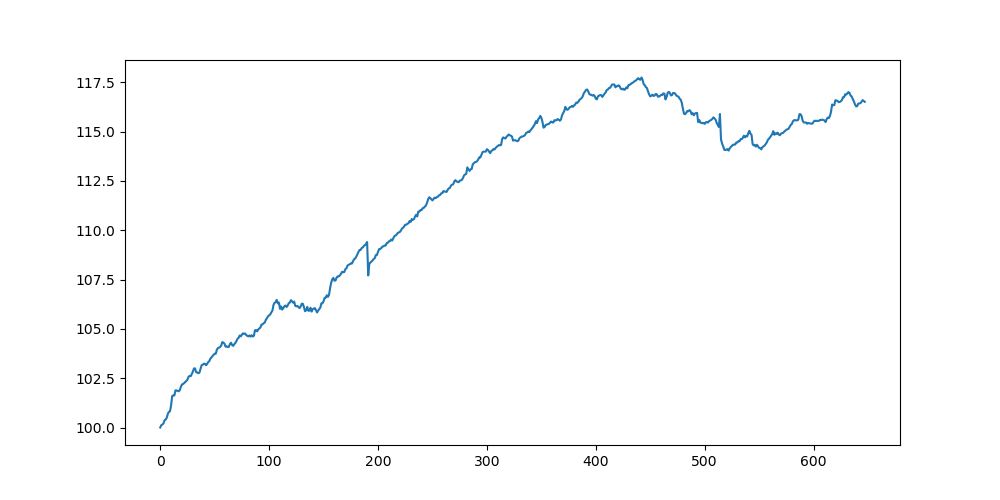}
    \caption{October Returns Example 2}
    \label{fig:my_label}
\end{figure}

\end{document}